\pdfoutput=1

\documentclass[11pt]{article}

\usepackage[final]{acl}

\usepackage{times}
\usepackage{latexsym}

\usepackage[T1]{fontenc}

\usepackage[utf8]{inputenc}

\usepackage{microtype}

\usepackage{inconsolata}

\usepackage{graphicx}

\title{Voice Interaction With Conversational AI Could Facilitate Thoughtful Reflection and Substantive Revision in Writing}

\author{
  Jiho Kim \\
  Calvin University \\
  \texttt{jihokim8@acm.org}
  \And
  Philippe Laban \\
  Microsoft Research \\
  \texttt{plaban@microsoft.com}
  \AND
  Xiang `Anthony' Chen \\
  University of California, Los Angeles \\
  \texttt{xac@ucla.edu}
  \And
  Kenneth C. Arnold \\
  Calvin University \\
  \texttt{kcarnold@alum.mit.edu}
}

\begin{document}
\maketitle

\begin{abstract}

Writing well requires not only expressing ideas but also refining them through revision, a process facilitated by reflection. Prior research suggests that feedback delivered through dialogues, such as those in writing center tutoring sessions, can help writers reflect more thoughtfully on their work compared to static feedback. Recent advancements in multi-modal large language models (LLMs) now offer new possibilities for supporting interactive and expressive voice-based reflection in writing. In particular, we propose that LLM-generated static feedback can be repurposed as conversation starters, allowing writers to seek clarification, request examples, and ask follow-up questions, thereby fostering deeper reflection on their writing. We argue that voice-based interaction can naturally facilitate this conversational exchange, encouraging writers' engagement with higher-order concerns, facilitating iterative refinement of their reflections, and reduce cognitive load compared to text-based interactions. To investigate these effects, we propose a formative study exploring how text vs. voice input influence writers' reflection and subsequent revisions. Findings from this study will inform the design of intelligent and interactive writing tools, offering insights into how voice-based interactions with LLM-powered conversational agents can support reflection and revision.

\end{abstract}

\section{Introduction}

Writing for effective communication requires more than just \textit{expressing} thoughts; it demands \textit{transforming} those thoughts to meet the expectations of an audience \citep{flowerWriterBasedProseCognitive1979b}. This transformation is achieved through \textit{revision}, which we define as any changes made to written content at any time, including both meaning-preserving edits and meaning-changing adjustments \citep{faigleyAnalyzingRevision1981a, fitzgeraldResearchRevisionWriting1987c}. Writing research suggests that substantive revision is facilitated through \textit{reflection} \citep{piankoReflectionCriticalComponent1979a}, in which writers critically examine their work from an external perspective to evaluate its effectiveness in addressing their rhetorical situation and fulfilling their communication goals \citep{flowerCognitionDiscoveryDefining1980a}.

However, writers often suffer from the curse of knowledge bias, which prevents them from reading their text as their audience would \citep{flowerDetectionDiagnosisStrategies1986b}. Consequently, feedback from others is a common way to facilitate reflection \citep{flowerWriterBasedProseCognitive1979b}. For example, many higher education institutions operate writing centers where tutors provide non-prescriptive and non-corrective feedback through conversational exchanges, encouraging writers to prioritize addressing higher-order concerns, such as thesis (or focus), audience engagement, organization, and content development, rather than lower-order concerns, such as grammar and syntax \citep{purdueowlHigherOrderConcerns, fitzgeraldOxfordGuideWriting2015, murphyStMartinsSourcebook2011, ryanBedfordGuideWriting2015}. This dialogue with the tutors helps writers gain critical distance from their work and make independent and substantive revisions.

Previous work has demonstrated user interface (UI) affordances that enable writers to use large language models (LLMs) to generate personalized and contextually adaptive feedback, questions, and advice, to facilitate reflection \citep{benharrakWriterDefinedAIPersonas2024b, kim2024FullAuthorshipAI}. However, these systems primarily support the generation of static feedback, questions, and advice, and lack the UI support for conversational exchanges characteristic of human tutoring. Yet, as highlighted by \citet{geroSocialDynamicsAI2023a} (in the second paragraph of Section 4.2.3), useful feedback comes from a back-and-forth conversational exchange, allowing the writers to clarify, ask follow-up questions, and refine their work based on the discussion. This discussion not only deepens the writer's understanding of feedback but also enables co-construction of meaning, which has been shown to improve feedback uptake and lead to substantive revisions \citep{zhaoFeedbackFeedbackOnFeedbackReFeedback2024}.

Furthermore, the input modality of interaction may significantly influence how people reflect on their writing through dialogue with LLM-powered conversational agents. A study by \citet{chalfonteExpressiveRichnessComparison1991a} showed that when co-authors gave feedback to each other on a collaborative writing task using spoken annotation, it imposed a lower cognitive load than written annotations, giving them more mental capacity to focus on higher-order concerns. However, there is limited evidence on how the modality of the writer's input, particularly spoken vs. written, impacts the quality of reflection when interacting with LLM-powered conversational agents.

Recent advancements in multi-modal LLMs highlight their ability to understand and interpret multi-modal instructions and generate outputs across different modalities, including text and speech \citep{zhang-etal-2023-speechgpt, wangBLSPBootstrappingLanguageSpeech2024, wang-etal-2024-blsp}. Commercially available multi-modal LLMs, such as GPT-4o \citep{openaiHelloGPT4o2024} and Gemini 2.0 Flash \citep{kavukcuogluGemini20Now2025}, demonstrate high accuracy and low latency in understanding and generating outputs across different modalities. This allows interaction designers to build both highly \textit{interactive} (i.e., responding quickly and appropriately) and \textit{expressive} (i.e., conveying emotions, social context, and nuanced meaning through non-verbal cues such as voice inflection, pitch, and tone) \citep{chalfonteExpressiveRichnessComparison1991a} conversational interfaces in ways that were not possible before. These advancements make our exploration of voice interactions with LLM-powered conversational agents timely.

We argue that the input modality, whether writers respond to feedback through voice or text, can influence the quality of their reflection. Specifically, we hypothesize that responding to feedback from an LLM-powered conversational agent using voice can (1) promote greater engagement with higher-order concerns, (2) facilitate more iterative refinement of one's reflections, and (3) reduce the cognitive load associated with reflection, compared to written input. To investigate this, we propose a formative study exploring how spoken versus written input affects writers' reflection and subsequent revision.

\section{Formative Study}

Consider a writer who has completed a rough draft of an argumentative essay and wants to revise it to better meet audience expectations. Before revising, they need to reflect on their work to set specific revision goals. They turn to reflection support systems, such as Textfocals \citep{kim2024FullAuthorshipAI} or Impressona \citep{benharrakWriterDefinedAIPersonas2024b}, seeking feedback, questions, and advice to help them reflect on their work and make decisions about what to revise. However, writers prefer feedback containing specific examples rather than vague comments. For example, one participant using Impressona noted that the system ``just tells me I have to do this and it doesn't come up with the specific examples'' (as said by P11 in \citet{benharrakWriterDefinedAIPersonas2024b}).

This illustrates a limitation in the aforementioned reflection support systems. As \citet{zhaoFeedbackFeedbackOnFeedbackReFeedback2024} and \citet{geroSocialDynamicsAI2023a} highlight, feedback is more useful when it is part of an interactive conversational exchange, allowing writers to seek clarification, request examples, and ask follow-up questions. This suggests an interaction design opportunity to repurpose static feedback, questions, and advice as conversation starters. Writers can use these starters to initiate reflective dialogues with LLM-powered conversational agents, to seek clarification, request examples, and ask follow-up questions, thus fostering thoughtful reflection.

We argue that the speech modality can naturally facilitate this conversational exchange. Prior research has shown that in collaborative writing, when co-authors provided feedback to each other using spoken annotations, the expressivity of spoken annotations encouraged a greater focus on higher-order concerns and promoted self-correction of the provided feedback compared to written annotations \citep{chalfonteExpressiveRichnessComparison1991a}. This motivates the following research questions:

\begin{description}
    \item[RQ1:] How might speaking to LLM-powered conversational agents, compared to typing, influence the depth and kinds of concerns writers reflect on in their work?
    \item[RQ2:] How does engaging in spoken conversations with LLM-powered conversational agents, as opposed to text-based interactions, shape the way writers refine and revisit their own reflections?
\end{description}

Furthermore, \citet{chalfonteExpressiveRichnessComparison1991a} suggested that the benefits of the speech modality might stem from spoken language production being less cognitively demanding than written language production \citep{bourdinWrittenLanguageProduction1994a}. This relative ease could allow writers to allocate more cognitive resources to their reflections. However, the cognitive demands of reflection itself, regardless of the modality, may also be significant. This raises the question:

\begin{description}
    \item[RQ3:] How do writers perceive the cognitive demands of speaking to LLM-powered conversational agents, compared to typing, and what factors influence these perceptions?
\end{description}

Finally, given the association between reflection and revision in writing \citep{piankoReflectionCriticalComponent1979a}, it is important to examine whether reflective dialogues with LLM-powered conversational agents lead to actionable changes in writing. This raises the question:

\begin{description}
    \item[RQ4:] How does reflecting with LLM-powered conversational agents influence the extent and depth of revisions in written content?
\end{description}

\subsection{Method and Measures}

To answer our research questions, we will employ a within-subjects experiment design, with participants counterbalanced across two conditions. All conditions will have interactive conversational capabilities, but will differ in the modalities provided: (1) written communication from both the user and the system, and (2) spoken input from the user with written output from the system.

Participants will be recruited using convenience sampling through multiple channels, including Calvin University's student, faculty, and staff community, as well as freelancing websites (e.g., Upwork), and crowdsourcing platforms (e.g., Prolific). We will seek individuals who can read, write, and speak English. Recruitment materials will provide information about the study and compensation\footnote{The first author's institutional review board approved the proposed study procedures.}.

Participants will be asked to write two rough drafts before the study, using argumentative writing prompts provided by the researchers. The length requirement of the rough drafts as well as the time the participant spends reflecting and revising in each conditions will be determined through a pilot study.

During the formative study, participants will use the formative system (see Section~\ref{sec:system_design}) modified for each condition to reflect on and revise their rough drafts. After completing each condition, participants will complete the NASA-TLX \citep{hartNASATaskLoadIndex2006} questionnaire to measure cognitive load. Following all conditions, a semi-structured interview will be conducted to gather qualitative insights on participants' experiences.

In terms of specific quantitative measures, we plan to examine the frequency of higher-order concerns (i.e., absolute count of higher-order concerns), after adjusting for the length of the response, as well as the proportion of higher-order concerns relative to the total number of concerns mentioned (i.e., relative count of higher-order concerns) in each condition. Coding of what qualifies as higher-order concerns or lower-order concerns will be based on criteria adapted from \citet{purdueowlHigherOrderConcerns}. For example, higher-order concerns include thesis or focus, audience and purpose, organization, and development, while lower-order concerns include sentence structure, grammar, and spelling. This criteria will be shared among the coders before data analysis to ensure inter-rater reliability. We will also record the number of conversational turns per minute as one proxy for engagement in conversation, and the time taken by participants to respond to the conversational agent as a proxy for cognitive processing. While we acknowledge that slower-speaking users or those processing complex responses may naturally take longer to reply, these measures will be interpreted with qualitative data (e.g., insights from aforementioned semi-structured interview) to better contextualize user behavior. Additionally, participants' revisions will be evaluated by experts using an argumentative essay rubric developed by \citet{ozfidanAssessmentStudentsArgumentative2022a}.

\subsection{System Design Considerations} \label{sec:system_design}

To ensure that our formative study isolates the effects of input modality, rather than introducing confounding factors such as interface novelty or feature differences, we design our formative system to align with established interaction patterns in existing LLM-powered writing tools. This helps maintain external validity and allows us to focus on how modality shapes writers' reflection and revision behaviors. In this section, we outline key interaction design considerations, centered on the dimensions of \textit{initiation}, \textit{contextualization}, and \textit{control}.

\subsubsection{Initiation}

In current LLM-powered conversational interfaces, users typically initiate interactions by explicitly asking questions or making requests. Reflection support systems, such as those proposed by \citet{benharrakWriterDefinedAIPersonas2024b} and \citet{kim2024FullAuthorshipAI}, follow a different approach. After users submit their drafts, these systems generate feedback, questions, or advice without requiring a specific prompt from the user, thereby initiating the interactions themselves. Our design explores the use of such system-initiated prompts (i.e., user-facing feedback, questions, or advice) as conversation starting points and incorporates UI affordances that enable writers to respond with clarifications, follow-up questions, or answers, either in text or speech, depending on the study condition.

\subsubsection{Contextualization}

Contextualization refers to how the writer's task is situated within the system to help writers maintain situational awareness of their work \citep{simkuteIroniesGenerativeAI2025}. Existing conversational interfaces, such as ChatGPT and Claude, achieve this by allowing users to open up a separate window (to the right of the conversational interface) called the Canvas \citep{openaiIntroducingCanvas2024} or Artifacts \citep{anthropicIntroducingClaude352024}. These UI affordances enable users to collaborate with LLM-powered conversational agents, allowing them to view, modify, and build on both their own work and LLM-generated content. Similarly, our design also explores providing a spatially distinct area for content creation and revision, helping users keep track of the context of their writing task.

\subsubsection{Control}

While contextualization helps in maintaining situational awareness, control is crucial for preserving the user's sense of ownership and agency over their content. Existing interfaces, such as those in ChatGPT and Claude, explicitly allow LLM-powered conversational agents to directly alter or expand upon the user's content. This capability, while useful for content generation and iteration, may compromise the user's sense of ownership and agency. In contrast, our design adopts an approach similar to previous work \citep{dangTextGenerationSupporting2022, benharrakWriterDefinedAIPersonas2024b, kim2024FullAuthorshipAI, labanChatExecutableVerifiable2024a} by supporting contextualization without allowing the LLM-powered conversational agents to directly modify or build upon the writer's content. This approach preserves the writer's control, and any feedback on writing (i.e., non-directive and non-prescriptive suggestions) from LLM is discussed with the writer, who reflects on them and decides whether to adopt the feedback by revising their content.

\section{Expected Contributions}

Through our formative study, we aim to inform the design of intelligent and interactive writing tools that support reflection through conversational exchanges with LLM-powered conversational agents. Specifically, we expect to make the following contributions: (1) providing evidence on the impact of speech modality in facilitating thoughtful reflection and (2) exploring the opportunity to transform static LLM-generated feedback, questions, and advice into dynamic conversational exchanges that encourage reflection and subsequent revision.

\section*{Acknowledgments}

We thank Jason Chew, Juyeong Kim, Heonjae Kwon, Ray Flanagan, Kor\'e Qualls, and the anonymous reviewers for their input and feedback. This work is supported by NSF CRII award 224614.

\bibliography{anthology, custom}
\end{document}